\documentclass[12pt,a4paper]{article}
\usepackage[cp1251]{inputenc}
\begin{document}
\large
\title{The interparticle interaction and noncommutativity
of conjugate operators in quantum mechanics II. Lightest nuclei.}
        \author{A.I.Steshenko \\ \\
\centerline{\it Bogolyubov Institute for Theoretical physics NASU,
Ukraine}  \\ }
\date{}
\maketitle

\begin{quotation}
A new model for calculating the structure of bound states of
interacting particles is considered. The model takes into account
the noncommutativity of the space and impulse operators plus the
correlation equations for the indeterminacy of these quantities.
The efficiency of the model is demonstrated by specific
calculations for some lightest nuclei.

\end{quotation}

\section{Introduction. The formulation of the model. }

\quad \,In our recent work ~\cite{1} we have proposed a new
quantum mechanical model for interacting bodies. The idea
~\cite{2}-~\cite{3}  that the coordinate and impulse operators for
different particles may be not commutative make up a basis of the
NOCE model (the NOCE model means the noncommutativity of the
operators and the correlation equations). Within  the framework of
the NOCE model, we have examined ~\cite{1} in detail the ground
and some excited states of Hydrogen-like ($H$-like) atoms. In the
present paper the $^2H\equiv D, ^3H$ and $^3He$ lightest nuclei
are studied by a given model.

In the case of $A$ identical particle the evident generalization
of the equations (5)-(8) from the paper ~\cite{1} yields the
following relations
\begin{equation}
\label{1}
    [x_k,\hat{p}_l^x]=[y_k,\hat{p}_l^y]=[z_k,\hat{p}_l^z]=\left\{
 \begin{array}{ll}
  i \hbar\, \beta, & \quad k=l; \\
  i\hbar\, \beta_o, & \quad k\neq l;
\end{array}\right.
\end{equation}
$$ \qquad \qquad \quad \mathbf{\hat{p}}_l= -i\hbar\,
\beta\cdot\mathbf{\bigtriangledown}_l,\qquad  k,l=1,2,...,A\, ;$$
where there are only two noncommutativity parameters  $\beta$ and
$\beta_o$ present. Assuming the commutator of the operator of the
particle coordinates and the operator of the total impulse
$\mathbf{P}$ to be equal $i\hbar$  one obtains from the above
mentioned equation a simple connection between $\beta$ and
$\beta_o$
\begin{equation}
\label{2}  1= \beta+(A-1)\,\beta_o\, .
\end{equation}
For this reason, only one additional equation is needed in order
to find the noncommutativity parameters $\beta$ and $\beta_o$. We
write this equation by examining the process of the measurement of
the coordinate of $k$-th particle with maximum accuracy.  The
reasoning similar to that employed in ~\cite{1} leads us to the
equation connecting the noncommutativity parameter $\beta_o$ and
the matrix element (ME) of the force $f_o\equiv \langle
|F|\rangle$, exactly,
\begin{equation}
\label{3}  \beta_o=\frac{2\hbar c}{\varepsilon^2} \gamma_o\cdot
f_o\, ; \qquad  \varepsilon \equiv m\,c^2  \, .
\end{equation}
Here $m$ is the nucleon mass, and $\gamma_o$ is one of the
correlation factors specified by the equations
\begin{equation}
\label{4} \begin{array}{ll} \Delta x_k\,\Delta p_l^x = \Delta y_k
\,\Delta p_l^y=\Delta z_k\,\Delta p_l^z=\left\{
   \begin{array}{ll}
  \frac{ \hbar }{2\gamma}\cdot \beta, & \quad k=l; \\ \\
  \frac{\hbar}{2\gamma_o}\cdot \beta_o, & \quad k\neq l;\\
   \end{array} \right.\\
     k,l=1,2,...,A.
   \end{array}
\end{equation}
Eqs.(2) and (3) yield one more useful relation :
\begin{equation}
\label{5}  \beta\,=\,1-\frac{2\hbar c}{\varepsilon^2} \cdot (A-1)
\gamma_o f_o\, .
\end{equation}
In the NOCE model, we have the analogy Schr$\ddot{o}$dinger
equation (SE)
\begin{equation}
\label{6}  \begin{array}{lcr} \left[-\frac{\hbar^2}{2m^{'}}
\sum\limits_{i=1}^A \nabla_i^2+\sum\limits_{i>j=1}^{A}
\hat{V}(\mathbf{r}_{ij})\right] \Psi_E(1,2,...,A)=E\,\Psi_E \, ,
\\\\  m^{'}=m/\left[1- \frac{2\hbar c}{\varepsilon^2} \cdot (A-1)
\gamma_{o} f_o\right]^2.
\end{array}
\end{equation}
Specific solutions to these equations can be found by the method
of successive iterations, where at the 1-st step the conventional
SE with the masses $m^{'}=m$, which the particles have in the
absence of the interaction ($\beta\equiv1$), has to be solved.
After that, on finding the wave function $\psi$, one can calculate
ME of the force $\langle |F|\rangle$ and the first value of the
commutation parameter $\beta$ distinct from unity. At the 2-nd
step, SE is solved with the modified particle masses
$m^{'}=m/\beta^2$. On finding the new $\psi$, we calculate the
quantity  $\langle |F|\rangle$ and compare it with the one
obtained at the 1st step. Then, we proceed with the iterations
until the values of the matrix element of the force $\langle
|F|\rangle$ obtained at subsequent steps will be virtually
indistinguishable. It is clear that, before starting the above
iteration process, we should specify the numerical value for the
correlation factor $\gamma_o$ entering Eqs.(6). To calculate its,
one can employ specific parameters of a given system based on
reliable experimental data. The way to practically implement this
will be described in detail hereinafter

\section{Deuteron and the variational Ritz principle.} \label{sect3}

\qquad Now we proceed to consideration of nuclear systems taking
as an example the simplest one, the deuteron. As is known, in
contrast to the atom theory, in the nuclear theory one has to deal
with a serious problem of choosing the nucleon-nucleon $(NN)$
forces. In view of this circumstance the accuracy of theoretical
estimates for a nucleus is considerably lower than that of similar
calculations in atomic physics. For this reason, taking into
account the calculations of $H$-atoms given in ~\cite{1}, such a
small correction in relation to the results of conventional SE may
evidence, at first sight, the inefficiency of application of the
NOCE model to nuclei. However, if it is remembered that the matrix
element (ME) of the force $\langle |F|\rangle$ for nucleon systems
should be several orders greater than similar values for $\langle
|F|\rangle$ in atoms, the contrary anxiety arises, meaning that
the application of the NOCE model to nuclear physics may result in
definitely unrealistic theoretical estimates.

Apparently, to clarify these questions, one needs to perform
specific numerical calculations. First, we can take as an example
the deuteron with a simple central potential in the Yukawa form
\begin{equation}
\label{7} V(r)=-V_o\frac{e^{-r/a}}{r/a}\, .
\end{equation}
The bound states of the deuteron are described by the wave
function $ \psi(\mathbf{r})=\\ \frac 1r
\chi_{nl}(r)Y_{ln}(\theta,\varphi)$ depending on the vector of the
relative motion $\mathbf{r}=\mathbf{r}_1-\mathbf{r}_2$. Let us
restrict ourselves to considering the ground $s$-state $(l=0)$,
then, the radial wave function $\chi_{n=1,l=0}(r)\equiv \chi(r)$
has to satisfy the simple equation
\begin{equation}
\label{8} \left[-\frac {\hbar^
2}{2\mu^{'}}\frac{d^2}{dr^2}+V(r)-E\right]\chi (r)=0\, ,
\end{equation}
where the reduced mass $\mu^{'}$ modified in the NOCE model equals
\begin{equation}
\label{9}
\mu^{'}=\frac{m_1^{'}m_2^{'}}{m_1^{'}+m_2^{'}}=\frac{m}{2\left(1-\frac{2\hbar
c}{\varepsilon^2}\gamma_o\, \langle |F|\rangle\right)^2};\quad
\varepsilon = mc^2\, .
\end{equation}
To write the equation (9), we use the relation
\begin{equation}
\label{10} m_{proton}=m_{neutron}\equiv m\, ,
\end{equation}
wich are natural for the case of identical particles. Satisfactory
solutions to the equation (8) can be found by means of the
variational Ritz principle. For the Yukawa-type potentials, good
results can be obtained by using the function
\begin{equation}
\label{11} \chi(r)=2 \eta^{3/2} r e^{- \eta\, r}\,
\end{equation}
as the variational one. Here the only variational parameter $\eta$
has to be found from the condition that the deuteron energy
\begin{equation}
\label{12} E(\eta)=\frac{\hbar^2}{2\mu^{'}}\,
\eta^2-V_o\,\frac{4(a\, \eta)^3}{(1+2a\,\eta)^2} \,
\end{equation}
must take extremum. In what follows, it is convenient, following
Ref.~\cite{4}, to introduce the notation
\begin{equation}
\label{13} K\equiv\frac{2\mu^{'}a^2}{\hbar^2}V_o\, ,\quad
p\equiv2a\,\eta \, .
\end{equation}
Then the above mentioned condition $dE/d\eta =0$ yields the cubic
equation with respect to the quantity $p$
\begin{equation}
\label{14}  p^3+(3-K)\,p^2+3(1-K)\,p+1=0\, .
\end{equation}
As is evidenced by the immediate calculation, the optimal value is
associated with the root $p_o$, which can be expressed via the
quantity $K$ introduced in (13) as
\begin{equation}
\label{15} p_o=\frac{K}{3}-1+2\sqrt{ \frac K3\left(1+\frac
K3\right)}\cdot\cos\left\{\frac \pi3-\frac 13
\arccos\left[\frac{K-\frac{K^2}{6}-\frac{K^3}{27}}{(\frac
K3+\frac{K^2}9)^{3/2}}\right]\right\}\, .
\end{equation}
The matrix element of the force
\begin{equation}
\label{16} F(r)=-\frac{d}{dr}V(r)=-a
\,V_o\left(\frac{1}{r^2}+\frac{1}{a r}\right)\,\exp(- \frac r a)\,
\end{equation}
on the functions (11) is equal
\begin{equation}
\label{17}
f_o\equiv\langle\chi(r)|F(r)|\chi(r)\rangle=8\frac{V_o}{a}(a\,\eta)^3
\frac{(1+a\,\eta)}{(1+2a\,\eta)^2}=\frac{V_o}{a}\cdot
p^3\frac{(1+\frac p2)}{(1+p)^2}\, .
\end{equation}
Substituting the value of the root $p_o$ from (61) instead of $p$
into the relation (17) and taking into account that the quantity
\begin{equation}
\label{18} K=\frac{a^2V_o}{(\hbar
c)^2}\cdot\frac{\varepsilon}{(1-\frac{2\hbar
c}{\varepsilon^2}\gamma_o\, f_o)^2}\,
\end{equation}
is, in turn, expressed  via $f_o$, we obtain some rather
complicated equation for ME of the force $f_o$. The equation found
in this way plays, apparently, the same role as the equation (33)
from ~\cite{1} does in the case of $H$-like atoms, namely, solving
this equation enables us to avoid the complicated (in
computational respect) procedure of successive iterations. The
necessary solutions have been found in the graphical way. In doing
so, from the set of all the solutions found (i.e., the points of
intersections of the straight line  $y=x\equiv f_o$\, with the
plot of the function \, $y=\varphi_2 (x)$\, representing the
right-hand side of the equation (17)) we selected the one
associated with the minimum value of $f_o>0$.

In the numerical calculations performed here, we used the
parameters of the potential (7) given in ~\cite{5}, $V_o=20.7$
MeV, $a=2.43$ fm. This potential was adjusted to fit the
experimental data on the scattering length and effective radius of
the triplet state in the "proton + neutron" system. The nucleon
mass was equal $\varepsilon\equiv m c^2=931.441$ MeV. The
calculations were carried out by using the various allowed $([0 <
\gamma_o \leq 1])$ values of the correlation factor $\gamma_o$.
The results obtained are given in the Table 1. In this table, the
ground state energy of the deuteron $E=E_D$, ME of the force
$f_o$, the optimum value of the root $p_o\equiv 2a\,\eta_o$, and
the values of the quantities
\begin{equation}
\label{19} 1-2\,x_o \equiv \beta ,\quad 2\,x_o\equiv \beta_o;\quad
x_o\equiv \gamma_o\, \frac{\hbar c }{\varepsilon^2}\, f_o\, ,
\end{equation}
defining the commutation relations (1) are presented as dependent
on the quantity $\gamma_o$.

As is seen from the table, the binding energy of the deuteron
$|E_D|$ (the experimental value $|E_D^{exp.}|=2.22457$ MeV) grows
with $\gamma_o$ from $|E_D|=2.2019$ MeV at $\gamma_o = 0.00001$ up
to $|E_D|=2.3826$ at $\gamma_o=1$. This means that in the NOCE
model under consideration, for the potential of $NN$-forces
chosen, the maximum possible increase of $|E_D|$ constitutes $\sim
8\%$ as compared to the calculation of $|E_D|$ based on the
conventional SE. With increasing the $\gamma_o$ parameter from
0.00001 to 1 the quantity $\beta$ diminishes insignificantly (by
less than 1\%) while the quantity $\beta_o$ grows from $\approx
0.906\cdot 10^{-7}$ at $\gamma_o=0.00001$ to $\beta_o \approx
0.01$ at $\gamma_o=1$.

The experimental binding energy of the deuteron corresponds to the
values $\gamma_o=0.13523$, $f_o=20.0812 MeV/fm$, and
$\beta=0.998\,765$, accompanied by the increase in the nucleon
mass by the value of $\triangle m= (\frac
{1}{\beta^2}-1)\,m=0.00247\,516 \,m$.

 \mbox{}
\begin{table} [1]
\caption{The properties of the ground state of the deuteron within
the NOCE model.}
 \begin{center}
\begin{tabular}{|cc|c|c|c|c|c|}
\hline \hline
 \multicolumn{2}{|c||}{$\gamma_o$ }
     & $E_D$; (MeV) & $f_o$; (MeV$/$fm) & $p_o=2a\,\eta$ & $\beta$& $\beta_o$  \\
\hline \multicolumn{2}{|c||}{0.00001           }
     &-2.20194412&19.9189032&2.27127197&0.99999991 &9.06148$\cdot 10^{-8}$  \\
\multicolumn{2}{|c||}{   0.01       }
     & -2.20360006  &19.9307771   &2.27190365   &0.99990934&9.06688$\cdot 10^{-5}$\\
\multicolumn{2}{|c||}{  0.10        }
                          & -2.21862818 & 20.038557 &2.27762869 & 0.99908847 &9.12591$\cdot 10^{-4}$\\
\multicolumn{2}{|c||}{  0.20    }
                          & -2.23559453 & 20.160273 & 2.28407581 & 0.99816586 &1.83426$\cdot 10^{-3}$ \\
\multicolumn{2}{|c||}{  0.30    }
                          & -2.25285174 & 20.284119 & 2.29061592 & 0.99723189 &2.76829$\cdot 10^{-3}$ \\
\multicolumn{2}{|c||}{  0.40    }
                          & -2.27040879 & 20.410160 & 2.29725172 & 0.99628626 &3.71398$\cdot 10^{-3}$ \\
\multicolumn{2}{|c||}{  0.50    }
                          & -2.28827498 & 20.538467 & 2.30398603 & 0.99532864 &4.671666$\cdot 10^{-3}$ \\
\multicolumn{2}{|c||}{  0.60    }
                          & -2.30646017 & 20.669115 & 2.31082180 & 0.99435870 &5.64166$\cdot 10^{-3}$ \\
\multicolumn{2}{|c||}{  0.70    }
                          & -2.32497459 & 20.802182 & 2.31776210 & 0.99337612 &6.62431$\cdot 10^{-3}$ \\
\multicolumn{2}{|c||}{  0.80    }
                          & -2.34382903 & 20.937744 & 2.32481016 & 0.99238052 &7.61997$\cdot 10^{-3}$ \\
\multicolumn{2}{|c||}{  0.90    }
                          & -2.36303480 & 21.075893 & 2.33196937 & 0.99137152 &8.62903$\cdot 10^{-3}$ \\
\multicolumn{2}{|c||}{  1.00    }
                          & -2.38260376 & 21.216715 & 2.33924329 & 0.99034874 &9.90348$\cdot 10^{-3}$\\
\hline \hline
\end{tabular}
\end{center}
\end{table}

\section{Mirror nuclei $^3H$ and $^3He$.} \label{sect4}

\qquad In this section, we consider the results of the
calculations of the ground states for the Tritium and Helium-3
nuclei.

In the case of a central $NN$-potential $V(r_{ij})$, the force $F$
entering the definition for the quantity $f_o$ is equal to the sum
of derivatives of $V(r_{ij})$ with respect to the relative
distance $r_{ij}$, i.e.
\begin{equation}
\label{20}  F=-\sum_{i>j=1}^A\frac{d}{dr_{ij}}V(r_{ij})\, ,\qquad
r_{ij}=|\mathbf{r}_i-\mathbf{r}_j|\,  .
\end{equation}
In specific calculations of light nuclei, one frequently restricts
consideration to the case of the central exchange potential
\begin{equation}
\label{21}  V(r_{ij})=-\sum_{S,T=0,1}V_{2S+1,2T+1}(r_{ij})\,
\hat{P}_{2S+1,2T+1}(ij)\, ,
\end{equation}
where the known projection operators \quad $\hat{P}_{2S+1,2T+1}$
\quad  do "cut" the relevant spin-isospin states of the
interacting nucleon $(ij)$-pair from the wave function \quad
$\Psi(1,2,...,A)$. The radial dependence of the components of
$NN$-potential  $V_{2S+1,2T+1}(r_{ij})$ can be presented, without
loss of generality, in the form of a series in Gaussian terms
\begin{equation}
\label{22}
V_{2S+1,2T+1}(r_{ij})=\sum_{\nu=1}^{\nu_{pot}}V_{2S+1,2T+1}^{[\nu]}\cdot
\exp\left(-\frac{r_{ij}^2}{\mu_\nu^2}\right) \, .
\end{equation}
Now we consider the problem of the bound states for the nuclei
$^3H$ and $^3He$. In the NOCE model, the conventional three-body
SE
\begin{equation}
\label{23} \hat{H}\,\Psi(1,2,3)=E\,\Psi,\quad
\hat{H}=\sum_{i=1}^3\left(-\frac{\hbar^2}{2m^{'}}\right)
\nabla_i^2+\frac{Z
e^2}{|\mathbf{r}_1-\mathbf{r}_2|}+\sum_{i>j=1}^3V(|\mathbf{r}_1-\mathbf{r}_2|)
\, ,
\end{equation}
with the modified nucleon mass
\begin{equation}
\label{24}  m^{'}=\frac{m}{\beta^2}, \qquad  \beta= 1 -
\frac{4\hbar c}{\varepsilon^2} \, \gamma_o\, f_o\, .
\end{equation}
is valid. The charge entering (23) is $Z=0$ for $^3H$ nucleus
(1-st and 2-nd particles represent neutrons), and $Z=1$ for the
nucleus $^3He$ ( 1-st and 2-nd particles are protons),
respectively.

In specific investigations of the structure of light nuclei, we
use as the $NN$-potential of the type (21)-(22) the two-Gaussian
Volkov potential ~\cite{6} with the following  parameters (I-st
version), 1-st component ($\nu=1$)\quad
$V_{31}^{[\nu=1]}=V_{13}^{[\nu=1]}=144.86$ MeV ,
$V_{33}^{[\nu=1]}=V_{11}^{[\nu=1]}=-28.972$ MeV, $\mu_1$=0.82 fm;
2-nd component
($\nu=2$)\quad$V_{31}^{[\nu=2]}=V_{13}^{[\nu=2]}=-83.34$ MeV,
$V_{33}^{[\nu=2]}=V_{11}^{[\nu=2]}=16.668$ MeV, $\mu_2$=1.60 fm.

In view of modern state of computer technology, it is convenient
to seek the solution of three-body SE by using the variational
Ritz principle with the wave function of the system being expanded
in a series  of the known basis functions. These latter may
contain one or another set of variational parameters, which,
actually, determines the flexibility of the basis chosen. In the
case of the lightest nuclei, satisfactory results were obtained by
using the simplest Gaussian-type functions. The efficiency of
using the Gaussian basis of functions in few-body problems has
been first realized by the theorists of the Moscow State
university in 1973-1975 (see, for instance, ~\cite{7} -
~\cite{10}). Really, even the first calculations performed within
the framework of the stochastic variational method (SVM) gave the
results of highest accuracy possible at that time. In view of
this, some authors sometimes refer to these calculations as the
"high precision" ones. Nowadays, the SVM method is widely and
successfully employed in theoretical studies of various quantum
systems consisting of small number of particles ~\cite{11} -
~\cite{13}.

That is why, it is reasonable to employ here SVM to calculate the
lightest nuclei under consideration. In doing so, the total wave
function $^3H$ and $^3He$ is presented, as usually, in the form of
a product of the space and spin-isospin function
\begin{equation}
\label{25}
\Psi(1,2,3)=\psi(\mathbf{r}_1,\mathbf{r}_2,\mathbf{r}_3)\cdot\chi(1,2,3)
\, .
\end{equation}
According to the experimental situation, the function $\chi$
represents the determinant
\begin{equation}
\label{26}
\chi(1,2,3)=\frac{1}{\sqrt{3!}}\left|\begin{array}{lcr}\chi_{\uparrow,
\uparrow}(1)&\chi_{\uparrow, \uparrow}(2)&\chi_{\uparrow,
\uparrow}(3)\\\chi_{\downarrow, \uparrow}(1)&\chi_{\downarrow,
\uparrow}(2)&\chi_{\downarrow, \uparrow}(3)\\\chi_{\uparrow,
\downarrow}(1)&\chi_{\uparrow,\downarrow}(2)&\chi_{\uparrow,
\downarrow}(3)\end{array}\right| \, ,
\end{equation}
composed by the one-particle spin-isospin functions
\begin{equation}
\label{27}
\chi_{\sigma\tau}(i)=\chi_{\frac12,m_\sigma}(i)\chi_{\frac12,
m_\tau}(i)\equiv\chi_{m_\sigma,m_\tau}(i) \, ,
\end{equation}
where the lower subscript $m_\sigma$ of the function (27) takes
two values denoted in (26) by an arrow "up" $\, \uparrow$ or
"down" $\, \downarrow$. Exactly same notation is used for the
other quantum number $m_\tau$. Resolving the determinant makes it
evident that the function $\chi(1,2,3)$ is antisymmetric with
respect to permutations of spin-isospin coordinates of 1-st and
2-nd particles. For this reason, by virtue of Pauli principle, the
$\psi$-function must be symmetric with respect to permutations of
space coordinates $\mathbf{r}_1$ and $\mathbf{r}_2$, which means,
in turn, that the expansion of $\psi$-function should be carried
out over the properly symmetrized Gaussian terms, i.e.,
\begin{equation}
\label{28}
\begin{array}{lcr}\psi_{L=0}(\mathbf{r}_1,\mathbf{r}_2,\mathbf{r}_3)=
\sum\limits_{j=1}^{j_{max}}C_j\cdot
u_j(\mathbf{r}_1,\mathbf{r}_2,\mathbf{r}_3)\, ,\\
\\ u_j(\mathbf{r}_1,\mathbf{r}_2,\mathbf{r}_3)=\hat{A}\,|j\rangle
\equiv\hat{A}\exp\{-\alpha_{12}^j\,r_{12}^2-\alpha_{13}^j\,r_{13}^2
-\alpha_{23}^j\,r_{23}^2\}=\\\\
=\sum\limits_{sym=1}^{2}\exp \left\{-\alpha_{12}^j(sym)\cdot r_{12}^2-
\alpha_{13}^j(sym)
\cdot r_{13}^2-\alpha_{23}^j(sym)\cdot r_{23}^2\right\}
\,.\end{array}
\end{equation}
The action of the symmetrization operator $\hat{A}$ on the usual
Gaussian term $|j\rangle$  gives rise to a sum consisting, in this
specific case, of two summands only with parameters
$\alpha_{kl}^j$  being replaced by $\alpha_{kl}^j(sym)$. The
result of the immediate examination of permutations associated
with the operator $\hat{A}$ is given in the Table 2.

As mentioned above, in numerical calculations we used the central
exchange Volkov potential, which contains the dependence on the
spin-isospin coordinates in the projection operators only. This
circumstance makes it possible to easily calculate the matrix
elements (ME) of the operator of $NN$-forces (21) with the
spin-isospin functions. In particular, for Volkov potential in the
case of the nuclei $^3H$ and $^3He$ under consideration, we obtain
\begin{equation}
\label{29} \begin{array}[t]{lcr} \langle \chi(1,2,3)\left |
\sum\limits_{i>j=1}^3 V(r_{ij})\right|\chi(1,2,3)\rangle =\\\\
 =V_{13}(r_{12})+\frac14 \left[ 3V_{31}(r_{13})+V_{11}(r_{13})+3V_{31}(r_{23})+
V_{11}(r_{23})\right] \, .
\end{array}
\end{equation}
Now, in order to write the system of linear equations for the
expansion coefficients of the wave function $C_j$, we calculate
the matrix elements of all the operators entering the definition
for the Hamiltonian (23) on the basis function
$u_j(\mathbf{r}_1,\mathbf{r}_2,\mathbf{r}_3)$. Earlier, the
analytical calculations of this kind have been performed in the
number of works (some of them are cited above). In view of this,
we cite here only selected results sticking to the notation of the
paper by N.N.Kolesnikov ~\cite{10}. Let us begin with the overlap
integral for the basis functions :
\begin{equation}
\label{30}   \langle u_{j^{'}}|u_j \rangle =
\sum\limits_{zym=1}^{2}\sum\limits_{sym=1}^2 \langle j^{'},zym|
j,sym \rangle;\qquad  \langle j^{'},zym| j,sym
\rangle=\left(\frac{\pi} {\sqrt{D^{[j^{'},j]}}}\right)^3 \, ,
\end{equation}
with the notation
\begin{equation}
\label{31} \begin{array}{ll} |j,sym \rangle
  \equiv \exp \left\{-\alpha_{12}^j(sym)\cdot r_{12}^2-\alpha_{13}^j(sym)
\cdot r_{13}^2-\alpha_{23}^j(sym)\cdot r_{23}^2\right\}\, ; \\\\
D^{[j^{'},j]}\equiv\alpha_{12}\alpha_{13}+\alpha_{12}\alpha_{23}+\alpha_{13}\alpha_{23};\quad
\alpha_{kl}\equiv \alpha_{kl}^j(sym)+\alpha_{kl}^{j^{'}}(zym)\, .
\end{array}
\end{equation}
It is clear that ME for any other operator on the functions (28)
represents a sum of the type (30), i.e, the analytical
calculations reduce, actually, to finding the partial ME with the
functions $|j,sym \rangle$. For the operator of the kinetic
energy, we obtain

\begin{table}
\caption{Symmetrized variational parameters
$\{\alpha_{kl}^{j}(sym)\}$ for the nuclei $^3H$ and $^3He$.}
\label{2}
\begin{center}
\begin{tabular}{|c||c|c|}                \hline
    $sym \qquad \rightarrow$     &   1   &  2  \\ \hline\hline
$\alpha_{12}^{j}(sym)  $ & \quad $\alpha_{12}^j$ &
\quad$\alpha_{13}^j$
\\ \hline
$\alpha_{13}^{j}(sym)  $ & \quad $\alpha_{13}^j$ & \quad
$\alpha_{12}^j$
\\ \hline
$\alpha_{23}^{j}(sym)  $ &  \quad$\alpha_{23}^j$ & \quad
$\alpha_{23}^j$
\\ \hline
\end{tabular}
\end{center}
\end{table}

\begin{table}
\caption{The properties of the ground states of the $^3H$ and
$^3He$ in the case $ \gamma_o=0.6$ and  the Volkov potential.}
\label{3}
\begin{center}
\begin{tabular}{|c||c|c|}                                   \hline
    $    $         &$   T\equiv ^3H $&$^3He$  \\ \hline\hline
$E^{exp.}; \, MeV        $ & $-8.482$     & $-7.718$
\\ \hline
$ E^{Q.mech.}; \, MeV$& $-8.464$     & $-7.759$
\\ \hline
$E^{theor.}; \, MeV$       & $-8.819$      & $-8.113$
\\ \hline \hline
$f_o; \, MeV/fm  $            &  $ 8.2483085 $& $8.4056275$
\\ \hline
$\beta $     & $0.995498$ & $0.995412$
\\ \hline
$\beta_o $     & $0.00225124$& $0.00229418$
\\ \hline
\end{tabular}
\end{center}
\end{table}

\begin{equation}
\label{32} \begin{array}{ll} \langle j^{'},zym|\hat{T}|j,sym
\rangle =\frac {3\hbar^2}{2 m^{'}} \frac{\langle
j^{'},zym|j,sym\rangle}{D^{[j^{'},j]}} \cdot \\\\
\cdot\sum\limits_{k=1}^{3}\sum\limits_{l^{'},l=1}^3\alpha_{kl}^{j^{'}}(zym)
\alpha_{kl}^j(sym)\left[
D_{l^{'}k}^{[j^{'}j]}+D_{lk}^{[j^{'}j]}-D_{l^{'}l}^{[j^{'}j]}\right];
\qquad (l^{'},l\neq k),
\end{array}
\end{equation}
with the notation ~\cite{10}
\begin{equation}
\label{33} \begin{array}{ll} D_{lk}^{[j^{'}j]}\equiv
\frac{\partial}{\partial {\alpha_{kl}}}D^{[j^{'}j]},\quad
D_{ll}^{[j^{'}j]}= 0;  \\\\
D_{12}^{[j^{'}j]}=\alpha_{13}+\alpha_{23}, \quad
D_{13}^{[j^{'}j]}=\alpha_{12}+\alpha_{23}, \quad
D_{23}^{[j^{'}j]}=\alpha_{12}+\alpha_{13}\, .
\end{array}
\end{equation}
The calculation of ME of the operator of potential energy is
associated with the expression
\begin{equation}
\label{34}  \langle j^{'},zym|V(r_{kl})|j,sym \rangle
=\sum\limits_{\nu=1}^{\nu_{pot}}V^{[\nu]}\left(\frac{\pi} {\sqrt
{D^{[j^{'}j]}+\frac{1}{\mu_{\nu}^2}D_{kl}^{[j^{'}j]}}}\right)^3\,
, \qquad (l^{'},l\neq k),
\end{equation}
where
\begin{equation}
\label{35} V(r_{kl})=\sum_{\nu=1}^{\nu_{pot}}V^{[\nu]}\cdot
\exp\left(-\frac{r_{kl}^2}{\mu_\nu^2}\right) \,
\end{equation}
is one of the components of the $NN$-potential (21) remained in
(29) after averaging over spin-isospin functions $\chi(1,2,3)$.
When considering the nucleus $^3He$, we have to calculate also the
Coulomb ME
\begin{equation}
\label{36}  \langle j^{'},zym\left |\frac{e^2}{r_{kl}}\right|j,sym
\rangle =\frac{2 e^2}{\sqrt {\pi}}\, \langle j^{'},zym|j,sym
\rangle \, \sqrt{\frac{D^{[j^{'}j]}}{D_{kl}^{[j^{'}j]}}}\, .
\end{equation}
Now, we have all the relations needed to write the system of
equations
\begin{equation}
\label{37}  H\cdot X = \lambda B \cdot X\, ,
\end{equation}
which is necessary to find the energy spectrum of the nucleus
$\{E_i\}=\lambda$ and the relevant coefficients in the expansion
$\{C_j\}=X$ of the space wave function
$\psi(\mathbf{r}_1,\mathbf{r}_2,\mathbf{r}_3)$. According to the
definition of the matrix elements $B\equiv||B_{j^{'}j}||$ and
$H\equiv||H_{j^{'}j}||$ in Eq.(37), we obtain
\begin{equation}
\label{38}   \begin{array}{ll} B_{j^{'}j}\equiv  \langle
u_{j^{'}}|u_j \rangle=\sum\limits_
{zym=1}^{2}\sum\limits_{sym=1}^2 \langle j^{'},zym|j,sym \rangle
\, ,\\\\H_{j^{'}j}= \sum\limits_{zym=1}^{2}\sum\limits_{sym=1}^2
\langle j^{'},zym | H=T+U+U_{Coul.}| j,sym \rangle \, .
\end{array}
\end{equation}
Thus, the energy of three nuclei is sought as a result of solving,
in the first place, the generalized problem for eigenvalues, and,
in the second place, the problem of finding the optimum values for
the variational parameters $\alpha_{kl}^j$ determining the basis
functions (28). In so doing, the most difficult procedure herewith
is the optimization. In performing numerical calculations, we
employed the known mathematical library $IMSL$, specifically, we
used the subroutine $DGVCSP$ ~\cite{14} in the generalized problem
for eigenvalues, and the subroutine $DBCONF$ ~\cite{15} for the
optimization procedure.

It should be kept in mind that the above mentioned calculations
are carried out only after the value of the nucleon mass
$m^{'}=m/\beta^2$ is found. The latter is determined by the
magnitude of ME of the force $f_o$ (see the relation (24)). In
this
\begin{equation}
\label{39} \begin{array}[t]{lcr} f_o = \langle \Psi(1,2,3)\left |
\sum\limits_{i>j=1}^3 -\frac{d
V(r_{ij})}{dr_{ij}}\right|\Psi(1,2,3)\rangle =\\\\=
\frac{1}{\langle
\Psi(1,2,3)|\Psi(1,2,3)\rangle}\sum\limits_{j^{'},j=1}^
{j_{max}}C_{j^{'}}C_j\sum\limits_{\nu=1}^{\nu_{pot}} \{
V_{13}^{[\nu]}\cdot G_{\nu}^{[j^{'},j]}(1,2)+\\\\
+\frac14(3V_{31}^{[\nu]}+V_{11}^{[\nu]})\cdot
G_{\nu}^{[j^{'},j]}(1,3) +
\frac14(3V_{31}^{[\nu]}+V_{11}^{[\nu]})\cdot
G_{\nu}^{[j^{'},j]}(2,3)\} \, ,
\end{array}
\end{equation}
with the partial ME being
\begin{equation}
\label{40}  \begin{array}[t]{lcr}  G_{\nu}^{[j^{'},j]}(k,l)\equiv
\langle j^{'},zym \left | \frac{2
r_{kl}}{\mu_{\nu}^2}e^{-\frac{r_{kl}^2}{\mu_{\nu}^2}} \right|j,sym
\rangle= \\\\ =\frac{4 \pi}{\mu_{\nu}^2}\left(\frac{\pi}
{D_{lk}^{[j^{'}j]}}\right)^{3/2}\left\{\frac{D_{lk}^{[j^{'}j]}}{D^{[j^{'},j]}
+\frac{1}{\mu^2} D_{lk}^{[j^{'}j]}}\right\}^2 \, .
\end{array}
\end{equation}
The quantities $D^{[j^{'},j]}$ and $D_{lk}^{[j^{'}j]}$ entering
this equation have been already specified by the relations (31)
and (33), respectively. On calculating ME of the force $f_o$ in
this way, we evaluate the mass $m^{'}$ by Eq.(24) and proceed to
solving SE with the modified mass $m^{'}$, i.e., we carry out the
process of successive iterations described above (Section I). In
so doing, we may put $f_o=0\quad (m^{'}=m)$ at the first step,
then, the 1-st iteration is identical to solving the conventional
SE.

Consider now the calculations of the nuclei $^3H$ and $^3He$
carried out for Volkov $NN$-potential ~\cite{6} with the model
parameter $\gamma_o=0,6$. In this calculation, the 15 functions
(28) were employed. As is seen from the definition (28), each of
these latter contains three independent variational parameters
$\alpha_{12}^j,\,\alpha_{13}^j$ and $\alpha_{23}^j$. Some of the
results of the calculations performed here are given in the Table
3. The values of the energy and ME of the force $f_o$ are given in
the units of MeV and MeV/fm, respectively. Since the
$NN$-potential used in calculations was adjusted to fit the basic
properties of light nuclei, the calculated binding energy
$E^{Q.mech.}$ turned out to be close to the experimental value. As
one would expect, the NOCE model overbinds the nuclei $^3H $ and
$^3He$ to some extent (see Table 3). The energy gain constitutes
$\sim 4\%$ as compared to $E^{Q.mech.}$, and the change in the
nucleon mass is  $\sim 0.9\%$.

It is clear that these numbers directly depend on a specific
choice of the $NN$-potential, as well as on the magnitude of the
parameter $\gamma_o$. For this reason, the theoretical results
presented here are qualitative in nature, in contrast to atomic
calculations ~\cite{1}. However, they are sufficient to make
estimates of the efficiency of the NOCE model under consideration.
In others words, the calculation performed made possible the
estimation of the order of the expected corrections to the basic
nuclear properties arising due to the allowance for the
noncommutativity of the coordinate and impulse operators of the
interacting particles.

\centerline{}

\end{document}